\documentclass[a4paper]{article}
\pdfoutput=1
\usepackage{graphicx}
\DeclareGraphicsExtensions{.png,.pdf}
\graphicspath{{./images/}}
\usepackage{float}
\usepackage{mathtools}
\usepackage{caption}
\usepackage{subcaption}
\usepackage{multirow}

\title{Higgs Measurements at a Muon Collider}
\author{
	Alexander Conway \\
	Department of Physics\\
	The University of Chicago
	and\\
	Hans Wenzel\\
	Computing Division\\
	Fermi National Accelerator Laboratory\\
	\\
	Faculty Advisor: Dr.\ Young-Kee Kim\\
	The University of Chicago}
\date{\today}

\begin{document}
\maketitle

\begin{abstract}
	In light of the recent discovery of an approximately 126 GeV Higgs boson at the LHC, the particle physics community is beginning to explore the possibilities for a next-generation Higgs factory particle accelerator. In this report we study the s-channel resonant Higgs boson production and Standard Model backgrounds at a proposed $\mu^+\mu^-$ collider Higgs factory operating at center-of-mass energy $\sqrt{s} = M_H$ with a beam width of 4.2 MeV. We study PYTHIA-generated Standard Model Higgs and background events at the generator level to identify and evaluate important channels for discovery and measurement of the Higgs mass, width, and branching ratios. We find that the $H^0\rightarrow b\bar{b}$ and $H^0\rightarrow WW^*$ channels are the most useful for locating the Higgs peak. With an integrated luminosity of $1\ fb^{-1}$ we can measure a 126 GeV Standard Model Higgs mass accurately to within 0.25 MeV and its total width to within 0.45 MeV. Our results demonstrate the value of the high Higgs cross section and narrow beam resolution potentially achievable at a muon collider.
\end{abstract}

\tableofcontents

\section{Introduction}
The Higgs field and its associated resonance, the Higgs boson were added to the Standard Model to solve the problem of how bosons and leptons acquire mass without breaking gauge invariance. The Higgs field and mechanism have since become relevant to fundamental theories about dark energy, dark matter and vacuum stability. The Higgs boson has been sought after for over forty years and until recently was the only remaining undiscovered fundamental particle predicted by the Standard Model.

In July 2012, the CMS and ATLAS experiments at the Large Hadron Collider in CERN announced the discovery of a new particle consistent with Standard Model predictions of a Higgs boson with a mass between 125 and 126 GeV. While further measurements have continued to indicate that the particle is a Higgs boson, it remains to be seen whether all of its properties are consistent with the Standard Model.\cite{cern-press} The LHC will not be able to make the precise measurements needed to test the Standard Model to its limits.\cite{lhc-higgs-width} The Standard Model predicts a width of $\Gamma_H = 4.21_{-3.8\%}^{+3.9\%} MeV$ for a Higgs particle with a mass of 126 GeV but the LHC will be limited to a mass measurement with uncertainty on the order of tens of MeV due to detector resolution.\cite{higgs-handbook}\cite{lhc-higgs-width} Some decay channels, such as $H^0\rightarrow b\bar{b}$ have very high irreducible physics backgrounds at the LHC, making branching fractions and partial widths very difficult to measure.\cite{H-meas-muller}\cite{lhc-higgs-width} 
	
	The fundamental importance of the Higgs boson and its exact properties motivates the construction of a new particle accelerator `Higgs factory'. There are many proposed candidates for a next-generation Higgs factory, including linear and circular $e^+e^-$ colliders, $\gamma\gamma$ colliders, plasma wakefield accelerators and a circular $\mu^+\mu^-$ collider.\cite{HF-tech-opt-delahaye} This report examines the potential ability of a proposed muon collider to fill that role and probe the Standard Model to its limits. We examine the physics backgrounds relevant to a muon collider operating at the Higgs s-channel resonance and explore an energy scanning search strategy for locating the narrow Higgs peak. We find that the high beam energy resolution and ability to use s-channel resonance Higgs production at a muon collider make it an attractive option for further research and development.

	In this report we assume a Standard Model Higgs. Unless stated otherwise, we assume a mass and width of 
\begin{equation}
	M_H = 126.0 GeV,\ \ \ \Gamma_H=4.21MeV \label{eq:higgs-props}
\end{equation}

\subsection{S-Channel Resonant Higgs Boson Production}
The Higgs boson's resonant production cross section is given by the Breit-Wigner formula. For a center of mass energy $\sqrt{\hat{s}}$, this is given by~\cite{han-higgs-measurement}
\begin{equation}
	\sigma(\mu^+\mu^- \rightarrow H^0) = \frac{4\pi \Gamma_H^2 Br(H^0\rightarrow \mu^+\mu^-)}{{(\hat{s} - M_H^2)}^2 + \Gamma_H^2 M_H^2}\label{eq:higgs-bw}
\end{equation}
$\Gamma_H$ is referred to as the `width' of the Higgs peak. The peak value of the cross section, using Standard Model values for the width and branching fractions of a 126 GeV Higgs, is approximately $64 pb$. The observable cross section is in practice the convolution of this Higgs peak with the energy distribution of the collider. We assume that the distribution of the center of mass energy is a Gaussian and unless otherwise stated, use a beam with a standard deviation in $\sqrt{\hat{s}}$ of 4.2 MeV, roughly the same as the Higgs peak. As will be shown later, this is an optimal width for discovering the Higgs. To calculate cross sections and to fit simulated data we numerically convolute the Higgs Breit-Wigner with a Gaussian. The peak value of the smeared cross section is $28.3 pb$.

\section{Muon Collider as a Higgs Factory}

	\subsection{Lepton Mass Coupling}
	The Higgs mechanism couples to the square of a leptons's mass so s-channel resonance Higgs production is enhanced by a factor of $4.28\times 10^4$ in a muon collider as compared to an electron collider\cite{mc-physics-eich}.
\begin{equation}
	\frac{g_{H\mu\mu}}{g_{Hee}} \propto \frac{m_{\mu}^2}{m_e^2} = 4.28\times 10^4~\cite{mc-physics-eich}
\label{lepton-coupling}
\end{equation}
In $e^+e^-$ colliders the only feasible channels for Higgs production are Higgs-strahlung ($e^+e^-\rightarrow ZH$) and vector boson fusion ($e^+e^-\rightarrow H\nu_e\bar{\nu_e}$), which have lower cross sections, higher physics backgrounds and do not allow for direct measurement of the Higgs mass and width.

	\subsection{Beam Energy Resolution}
Many properties of a muon collider make it an attractive option for a Higgs factory. The high mass of the muon compared to the electron, a ratio of about 200, suppresses radiative effects like beamstrahlung, allowing for a much greater beam energy resolution than an electron-positron collider. Estimates for achievable beam energy resolution at center of mass energies near 126 GeV are on the order of 1 to 10 MeV, surpassing the capabilities of any $e^+e^-$ machine by a factor of ten or greater.\cite{map-palmer} This high energy resolution allows for direct measurement of the Higgs width by `scanning' the beam energy across the s-channel resonance peak. 

	With a beam energy resolution equal to the Higgs total width, the Higgs production cross section is $\sigma (\mu^+\mu^-\rightarrow H^0) = 28.3 pb$.\cite{mc-physics-eich} Thus it is feasible to directly measure the Higgs width by `scanning' the beam energy across the Higgs peak. Estimates for the luminosity of a 63 GeV per beam muon collider are on the order of $L = 10^{32} cm^{-2}s^{-1}$, suggesting the production of about 50,000 to 100,000 Higgs bosons per year when taking data at the s-channel resonance. This would be competitive with a 126 GeV per beam linear $e^+e^-$ collider with $L = 2.0\times 10^{34} cm^{-2}s^{-1}$ producing Higgs bosons in the Higgs-strahlung and vector boson fusion processes.\cite{map-palmer}

\section{Machine Induced Background}
\begin{figure}
	\includegraphics[width=\textwidth]{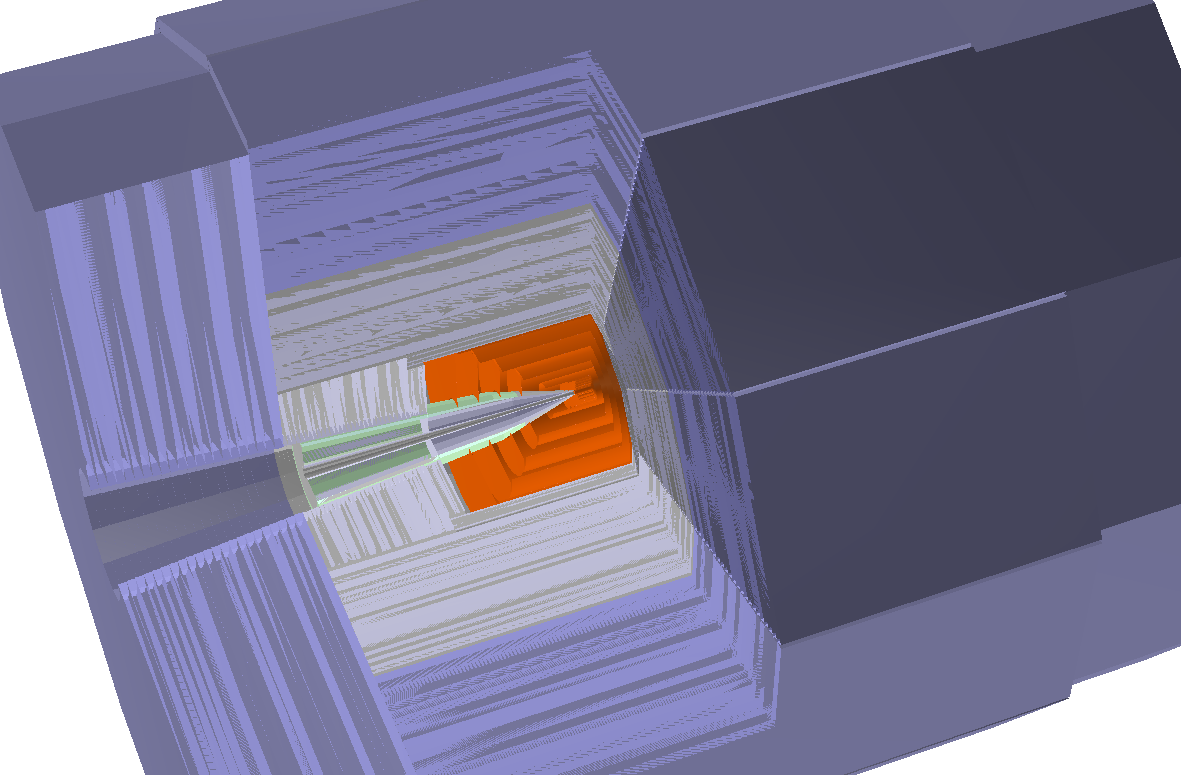}
	\caption{Conceptual rendering of MCDRCal01 detector. The silicon tracker (orange) outer radius is $122$ cm. The electromagnetic and hadronic calorimeters (gray) are made of Bismuth Germanium Oxide, or BGO.\@ The EM calorimeter has an inner radius of 125 cm and has ten $2 cm$ thick layers, each segmented by a $1 cm \times 1 cm$ grid. The Hadronic calorimeter has an inner radius of $146 cm$ and has thirty$ 5 cm$ thick layers segmented by a $2 cm \times 2 cm$ grid. The muon calorimeter is made of steel-235, has an inner radius of 300 cm and has 22 $10 cm$ thick layers segmented by a $10 cm \times 10 cm$ grid. The detector has a 5 Tesla magnetic field. All calorimeters are fully sensitive. The cones are made of Tungsten surrounding the beamline and form an angle of 15 degrees. See Appendix~\ref{sec:evt-display} for Higgs events simulated using this detector model.\label{fig:detector}}
\end{figure}

One disadvantage of a muon collider is that a large number of muons decay in the beamline, causing significant background. While we are mostly ignoring this for the purposes of this report, we do include a $\cos(\theta)$ restriction on our particles corresponding to a cone angle of 20 degrees. This is a conservative estimate motivated by the need for the cone to be thick enough to provide some shielding. Figure~\ref{fig:detector} is a conceptual rendering of MCDRCal01, a detector with a dual-readout capability that measures both ionizing radiation and Cerenkov photons in the totally active calorimeters. This detector will be used in future simulation studies but is not used as a reference for this report. Simulated Higgs event displays are provided in Appendix~\ref{sec:evt-display}.

\section{Physics Background}
\begin{figure}[h]
	\includegraphics[width=0.9\textwidth]{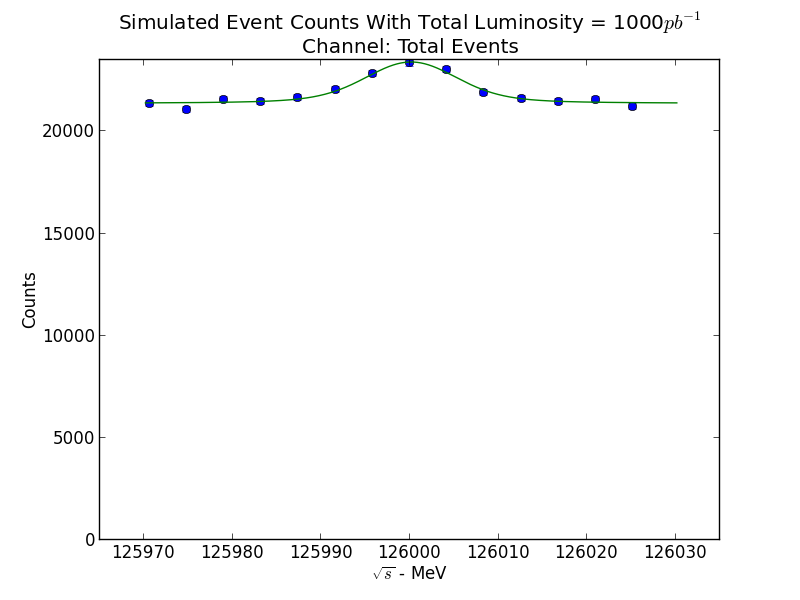}
	\caption{Simulated event counts for a scan across a 126.0 GeV Higgs peak with a 4.2 MeV wide Gaussian beam spread, counting all events except for $Z^0\rightarrow \nu_{\ell}\bar{\nu_{\ell}}$ decays. Data is taken in a 60 MeV range centered on the Higgs mass in bins separated by the beam width of 4.2 MeV. Total luminosity is $1~fb^{-1}$. Event counts are calculated as Poisson-distributed random variables and the data is fit to a Breit-Wigner convoluted with a Gaussian peak plus linear background. Fitted values of the free parameters are in Table~\ref{table:m-g-meas}.}
\label{data-fit-total-raw}
\end{figure}

The most significant background for s-channel resonance Higgs production at a muon collider is the production of Z bosons. The Higgs cross section, smeared by a 4.2 MeV beam is 28.3 pb. The cross section of the Z background is 376 pb, but 20.04\% of these Z's decay into pairs of neutrinos and a photon, bringing the cross section to 301.4 pb and $S/\sqrt{B}$ to 1.63. This cross section remains essentially flat in the region around the Higgs peak and will be treated as such in this report. Figure~\ref{data-fit-total-raw} shows simulated data of a scan across a 126.0 GeV Higgs peak counting all events except for $Z^0\rightarrow \nu_{\ell}\bar{\nu_{\ell}}$. The data is fitted to a Breit-Wigner convoluted with a Gaussian with three free parameters; $\Gamma_H$, $M_H$ and $Br(H^0\rightarrow X)$. The fixed parameters are the background cross section $\sigma(Z^0\rightarrow X)$, the beam width $\sigma_{beam}$ and the total integrated luminosity $\mathcal{L}$. The fit gives a width of $4.56\pm 1.52$ MeV, an error in the mass measurement of $0.13\pm0.16$ MeV and a branching ratio of $0.96\pm0.04$. 

	\subsection{Low-Mass Z bosons}
\begin{figure}[h]
	\centering
	\includegraphics[width=\textwidth]{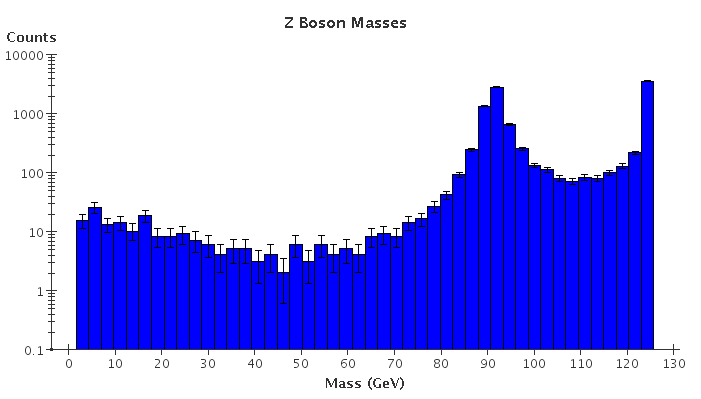}
	\caption{Z boson masses in 10,000 PYTHIA-simulated $\mu^+\mu^-\rightarrow Z$ events at $\sqrt{s}=125.0GeV$. The low-mass region is dominated by the Drell-Yan process. There is a peak around the Z mass where intial-state Bremsstrahlung radiation allows the creation of an on-shell Z. The third region of interest is the peak at $125GeV$, the center of mass energy. This represents a process with no initial state radiation where the off-shell Z's produced are indistinguishable from the Higgs.}
\label{z-mass-plot}
\end{figure}

\begin{figure}[h]
	\centering
	\begin{subfigure}[b]{0.4\textwidth}
		\centering
		\includegraphics[width=\textwidth]{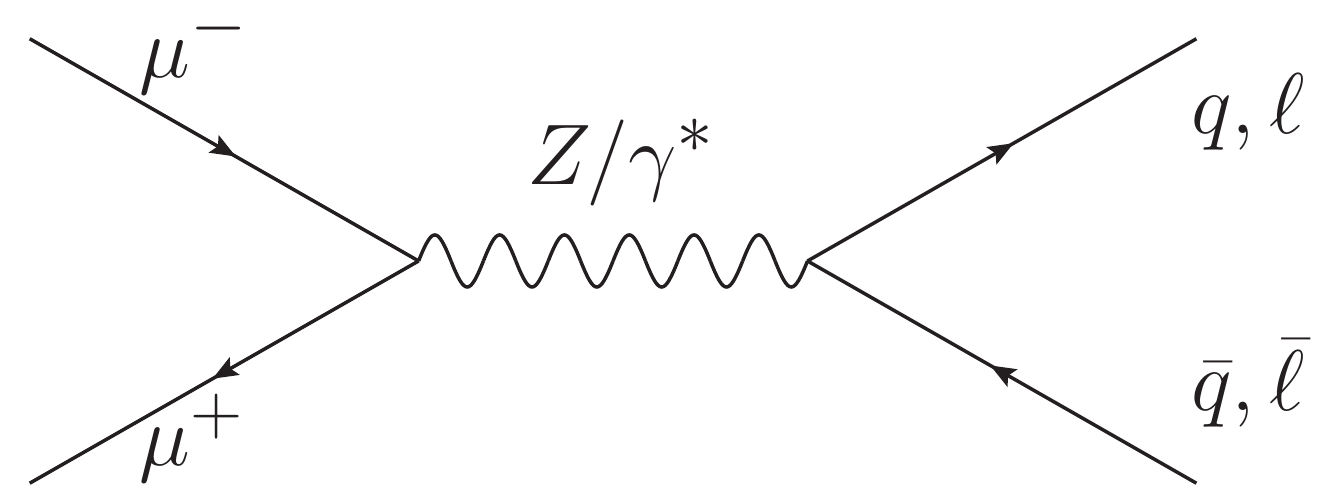}
		\caption{Irreducible background: $\mu^+\mu^-\rightarrow Z/\gamma^*$ with $M_{Z^*} = \sqrt{s}$.}
\label{fig:mumu-zstar-ql}
	\end{subfigure}
	\qquad
	\begin{subfigure}[b]{0.4\textwidth}
		\centering
		\includegraphics[width=\textwidth]{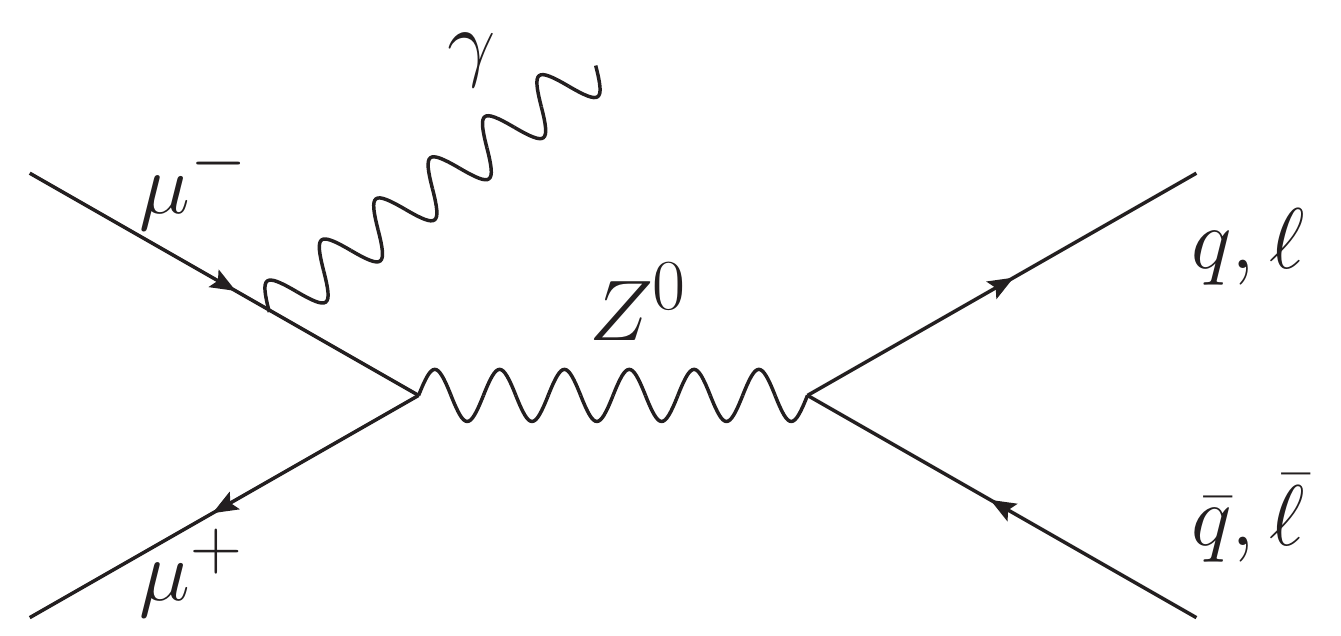}
		\caption{Reducible background: $\mu^+\mu^-\rightarrow Z^0,\gamma$ with $M_{Z^0} < M_{H^0}$.}
\label{fig:mumu-zg-ql}
	\end{subfigure}
	\caption{Standard Model backgrounds at a $\mu^+\mu^-$ collider operating at $\sqrt{s}=126\ GeV$}
\label{sm-bkg-feyn}
\end{figure}

	Fortunately, this background is reducible. The s-channel resonance production of Higgs bosons only happens with a center of mass energy within a few MeV of its peak. Z bosons however are produced in several different processes with a wide range of masses, as seen in Figure~\ref{z-mass-plot}. At an s-channel Higgs factory muon collider, Z bosons are primarily produced as real, on-shell bosons along with an intial state photon that makes up the difference in energy between the Higgs s-channel and the Z mass (Fig.~\ref{fig:mumu-zg-ql}). There is also a small number of very low mass Z bosons produced in a Drell-Yan process. The only events that are theoretically indistinguishable from Higgs events are those where a virtual $Z$ is produced at the center of mass energy and decays into a channel shared with the Higgs (Fig.~\ref{fig:mumu-zstar-ql}).

\begin{figure}[h]
	\includegraphics[width=0.9\textwidth]{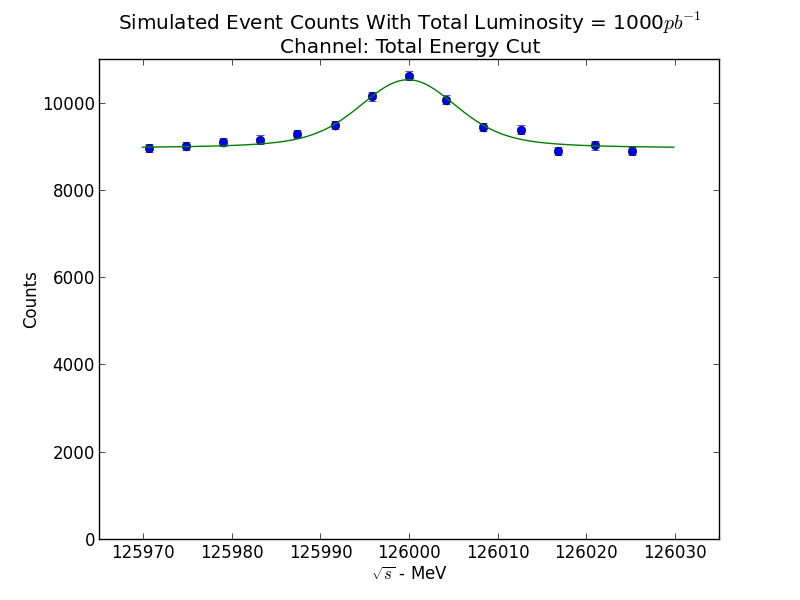}
	\caption{Simulated event counts for a scan across a 126.0 GeV Higgs peak with a 4.2 MeV wide Gaussian beam spread, counting all events with a total energy of at least 98.0 GeV visible to the detector. Data is taken in a 60 MeV range centered on the Higgs mass in bins separated by the beam width of 4.2 MeV. Event counts are calculated as Poisson-distributed random variables and the data is fit to a Gaussian peak plus linear background. The fit width is $5.16\pm0.24$ MeV and the error in the mass measurement is $0.26\pm0.19$ MeV.}
\label{data-fit-total-cut}
\end{figure}

Before looking into how the kinematics of these events might differ from Higgs events, the simple thing to do is a cut on the total energy potentially visible to the detector. This is accomplished by summing the energies of all final state particles which pass a $\cos{\theta} < 0.94$ cut and finding the energy cut which maximizes $S/\sqrt{B}$. The $\cos{\theta}$ cut is effective because most of the high-energy initial state radiation is colinear with the beam. We use a cut of $E_{total} > 98.0\ GeV$, which selects 79.2\% of the Higgs signal events and 41.9\% of the Z background. This results in an effective Higgs cross section of 22.4 pb and a background of 126.4 pb, bringing $S/\sqrt{B}$ to 1.99. Figure~\ref{data-fit-total-cut} shows simulated data using these results, with a fitted width of $5.57\pm 1.33$ MeV and an error in the mass measurement of $-0.02\pm0.14$ MeV. This simple cut has already proven to be a marginal improvement but there is much more that can be done by focusing on individual decay channels.

\begin{table}
	\begin{center}
		\begin{tabular}{|l|l|l|l|l|l|}
			\hline
			\multirow{2}{*}{Decay Mode} & \multicolumn{2}{|c|}{$Z$} & \multicolumn{2}{|c|}{$H^0$} \\ \cline{2-5}
			& BR & $\sigma$ (pb) & BR & $\sigma$ (pb) \\ 
			\hline
			$u\bar{u}$,$d\bar{d}$,$s\bar{s}$ & 0.427 & 160.6 & 0.0003 & 0.009 \\ \hline
			$c\bar{c}$      & 0.119 & 44.8 & 0.032 & 0.91 \\ \hline
			$b\bar{b}$      & 0.152 & 57.2 & 0.584 & 16.5 \\ \hline
			$e^+e^-$        & 0.034 & 12.8 & --- & --- \\ \hline
			$\mu^+\mu^-$    & 0.034 & 12.8 & --- & --- \\ \hline
			$\tau^+\tau^-$  & 0.034 & 12.8 & 0.071 & 2.01 \\ \hline
			$\nu_{\ell}\bar{\nu_{\ell}}$    & 0.200 & 75.4  & --- & --- \\ \hline
			$gg$                    & ---   & ---   & 0.053 & 1.50 \\ \hline
			$\gamma\gamma$  & ---   & ---   & 0.003 & 0.085 \\ \hline
			$WW^*$                  & ---   & ---   & 0.226 & 6.39 \\ \hline
			$Z^0Z^0$                & ---   & ---   & 0.028 & 0.79 \\
			\hline \hline
			Total:  & 1.0 & 376.3 & 1.0 & 28.3 \\ \hline
		\end{tabular}
	\end{center}
	\caption{Branching fractions and effective cross sections for Standard Model decay modes of Higgs and Z bosons. Higgs cross sections are calculated as the peak value of the Higgs peak Breit-Wigner convoluted with a Gaussian of width 4.2 MeV to simulate the effect of beam smearing. Branching fractions are taken from PYTHIA 6.4 event generation output.\label{bfs-xsects}}
\end{table}

	\subsection{$b\bar{b}$}
	Table~\ref{bfs-xsects} compares the branching ratios and cross sections of the Z background with the Higgs signal. The largest Higgs decay channel is $H^0\rightarrow b\bar{b}$, which makes up 58\% of Higgs decays at this mass, a branching fraction proportionally large to $Br(Z^0\rightarrow b\bar{b}) = 15.2\%$. We assume a b-tagging efficiency and purity of 1, so the cross sections for the decays are 16.5 and 57.2 pb, respectively. The fitted values for the mass, width and branching ratio of the Higgs using b-tagging are shown in Table~\ref{table:m-g-meas} and a fit to simulated data can be found in Appendix~\ref{sec:sim-evt-cts}.

	In both signal and background the visible energy spectrum is very similar to the spectrum of the combined channels, so the same total energy cut of $E_{tot}>98.0 GeV$ maximizes $S/\sqrt{B}$. Cuts on the event shape, the magnitude of the thrust and major axis, can further enhance the signal. The event shape is calculated by finding the axis which maximizes the sum of all particle momenta projected onto a single axis, called the `thrust axis'. This is then repeated for an axis perpendicular to the first and then a third orthogonal to both. The thrust is the normalized sum of the projection of all particle momenta against the thrust axis and the major axis value is the normalized sum of the projections against the secondary axis. Because the Higgs is never created in events with significant beamstrahlung it is always produced with low momentum. Z bosons produced with mass lower than the beam center-of-mass energy are `boosted' by the beamstrahlung photon. This boost lowers the thrust and raises the major axis values, so it is a useful indicator for channels with particular event shape profiles.
	
\begin{figure}
	\includegraphics[width=\textwidth]{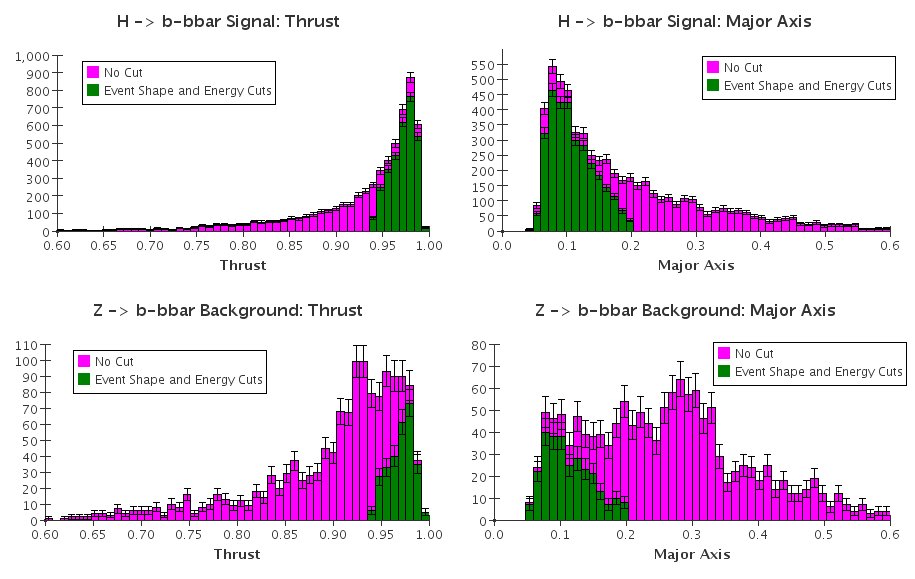}
	\caption{Effects of event shape and energy cuts on Higgs $b\bar{b}$ signal and background. Cuts were made by selecting events with total energy $E_{tot} > 98.0GeV$ visible to the detector, thrust between 0.94 and 1.0 and major axis between 0.0 and 0.2. The signal is reduced to 52\% and the background to 15\%.}
\label{bbbar-en-thrust-cuts}
\end{figure}

\begin{figure}
	\includegraphics[width=\textwidth]{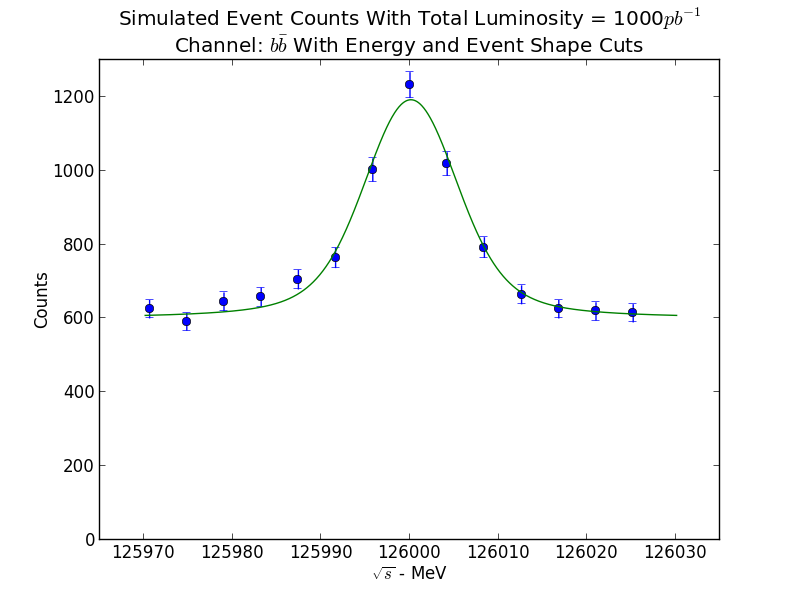}
	\caption{Simulated event counts for a scan across a 126.0 GeV Higgs peak with a 4.2 MeV wide Gaussian beam spread, counting $X\rightarrow b\bar{b}$ events with a total energy of at least 98.0 GeV visible to the detector and cutting on event shape parameters. Data is taken in a 60 MeV range centered on the Higgs mass in bins separated by the beam width of 4.2 MeV. Event counts are calculated as Poisson-distributed random variables and the data is fit to a Breit-Wigner convoluted with a Gaussian plus linear background. The fit width is $4.78\pm0.48$ MeV, the error in the mass measurement is $0.01\pm0.05$ MeV and the branching ratio is measured at $0.271\pm0.001$. Total luminosity is $1000pb^{-1}$, or $71.4 pb^{-1}$ per point.}
\label{data-fit-bbbar-cut}
\end{figure}

Figure~\ref{bbbar-en-thrust-cuts} shows the signal and background thrust and major axes before and after cutting on the total energy and event shape values. The cuts were made by selecting events with $E_{tot} > 98.0 GeV$, thrust values between 0.94 and 1.0 and major axis values between 0.0 and 0.20. We continue to assume perfect b-tagging. These cuts reduce the $b\bar{b}$ signal by 52\% and the background by 15\%, bringing the effective cross sections to 8.64 and 8.45 pb respectively. This brings the $S/\sqrt{B}$ ratio to 2.97, a dramatic improvement over simple energy cuts or b-tagging alone. Figure~\ref{data-fit-bbbar-cut} shows a simulated scan of the Higgs peak with a fit to a Breit-Wigner convoluted with a Gaussian.

\subsection{$H^0\rightarrow WW^*$}
There are several channels with very little physics background that are of importance, despite their smaller cross sections. One of these is the $H^0\rightarrow WW^*$ decay mode, with a branching fraction of 0.226 (cross section 6.39 pb) and no real background from the corresponding Z decays. The W boson decays into a charged lepton and corresponding neutrino 32.4\% of the time, with effectively equal rates for each type of lepton. The majority of the remaining branching fraction is the decay into pairs of light quarks. While it is certainly possible to reconstruct W bosons from four-jet events, in this report we focus on the decays with missing energy in the form of neutrinos since they can be identified by the presence of one or two isolated leptons and missing energy and are the most common. Further study will be required for a detailed analysis of the four-jet case. Since the W boson decays into a lepton and neutrino 32.4\% of the time and we require at least one such decay between a pair of W's, these make up 54.3\% of $WW^*$ events. Thus the theoretical cross section is 6.39 pb with virtually no background.

Because the detector will have a non-sensitive cone, there will be a small amount of `fake' background, eg.~when the photon in the decay $\mu^+\mu^-\rightarrow Z^0+\gamma\rightarrow\ell^++\ell^-$ boosts the two leptons and disappears into the cone as missing energy. Figure~\ref{fig:evt-disp-ww-e-tau-nunu} in Appendix~\ref{sec:evt-display} shows an example event display for a $WW^*$ decay into two leptons and illustrates the characteristic missing energy of these events. It is difficult to estimate the true background from processes such as these, but given the low branching ratios of $Z^0$ to lepton pairs and the kinematic and geometric constraints for `fake' background, it is safe to assume that the background will be fairly low in this channel. Therefore we use the rate assumed by Han \emph{et al}\cite{han-higgs-measurement}, a cross-section of 0.051 pb. Plots of simulated data for the $WW^*$ channel can be found in Appendix~\ref{sec:sim-evt-cts} and fitted values in Table~\ref{table:h-fits}.

\subsection{$\tau^+\tau^-$}
The $\tau^+\tau^-$ channel is dominated by the background, but the Higgs branching ratio of 0.071 is not insignificant. The $Z^0\rightarrow \tau^+\tau^-$ process has a branching ratio of 0.034, giving it an effective cross section of 12.8 pb, compared to the 2.01 pb cross section for the Higgs. However, the boost given to the lower mass Z bosons means the background can be further distinguished using total energy and event shape parameters. 

\begin{figure}
	\includegraphics[width=\textwidth]{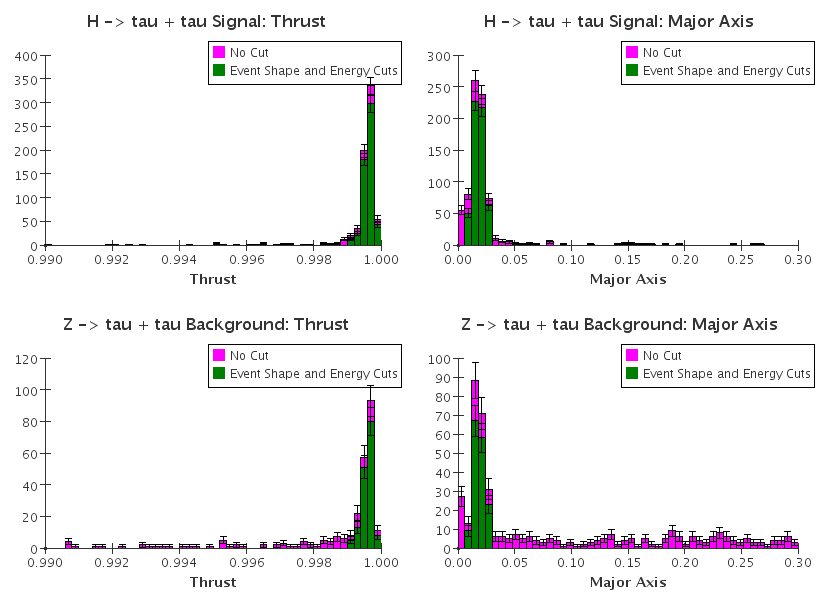}
	\caption{Effects of event shape and energy cuts on Higgs $\tau^+\tau^-$ signal and background. Cuts were made by selecting events with total energy $E_{tot} > 60.0GeV$ visible to the detector, thrust between 0.999 and 1.0 and major axis between 0.07 and 0.032. The signal is reduced to 78\% and the background to 39\%.}
\label{tautau-en-thrust-cuts}
\end{figure}

The $\tau$ is a short-lived particle and every $\tau$ decay channel involves the production of a $\tau$ neutrino. This makes the total visible energy less useful as a cut parameter than it was for $b\bar{b}$, since there are random amounts of missing energy. We require at least 60.0 GeV to be visible because background dominates below this value due to boosted Z's. Event shape parameters, however, are very useful here since $\tau$ decays typically do not create a widespread shower. We require the thrust to be between 0.999 and 1.0 and the major axis to be between 0.007 and 0.03. This cut reduces the signal to 78\% of its original value and the background to 39\%, bringing the Higgs cross section to 1.58 pb and the background to 4.97 pb, as seen in Figure~\ref{tautau-en-thrust-cuts}. The cut is specific enough that it is not necessary to assume anything else about the events, such as a perfect $\tau^+\tau^-$ tag. Fewer than 0.2\% of the Higgs decays that pass the cut are not $\tau^+\tau^-$ events and only 6.4\% of the background events that pass are misidentified. The effective background cross section above is calculated from all the events which pass the cut. Plots of simulated data can be found in Appendix~\ref{sec:sim-evt-cts} and fitted values can be found in Table~\ref{table:h-fits}.

\subsection{$H^0\rightarrow \gamma\gamma$}
The final channel examined in this report is the $H^0\rightarrow \gamma\gamma$ channel. The Higgs branching fraction for this channel is only 0.3\%, but the events can't be easily identified by selecting events with two photons with equal energy adding up to $\sqrt{s}$ and high thrust. About 10\% of these events are lost when one or both photons hit the cone and there is no background so the cross section is 0.077 pb. The high purity of this channel is a great advantage, but the small cross section makes it impractical for scanning the beam energy to find the Higgs peak as it takes a great deal of luminosity to expect more than a few events on the peak. This channel will require much luminosity but may prove very useful for precise measurements of the Higgs.

\begin{table}
	\centerline{
	\begin{tabular}{|c|l|l|l|l|l|l|l|}
		\hline
		\multirow{2}{*}{Channel} & \multicolumn{3}{|c|}{$\mu^+\mu^-\rightarrow H^0 \rightarrow X$} & \multicolumn{3}{|c|}{$\mu^+\mu^-\rightarrow Z\gamma^* \rightarrow X$} & \multirow{2}{*}{$S/\sqrt{B}$} \\
		\cline{2-7}
		& Br &\multicolumn{2}{|c|}{$\sigma$ (pb)} & Br & \multicolumn{2}{|c|}{$\sigma$ (pb)} &  \\
		\hline
		 \multirow{2}{*}{Total}
			& \multirow{2}{*}{1.0}	& $\sigma_s$		& 28.3	
			& \multirow{2}{*}{1.0}	& $\sigma_b$		& 301.4	& 1.63	\\
			&&$\sigma_{eff}$ & 22.4 && $\sigma_{eff}$& 126.4	& 1.99	 \\
		\hline
		\multirow{2}{*}{$b\bar{b}$}	
			& \multirow{2}{*}{0.584}	& $\sigma_s$	& 16.5	
			& \multirow{2}{*}{0.152}	& $\sigma_b$	& 57.2	& 2.18	 \\
			&&$\sigma_{eff}$ & 8.64	&	& $\sigma_{eff}$& 8.45	& 2.97	 \\
		\hline
		\multirow{2}{*}{$WW^*$}	
			& \multirow{2}{*}{0.226}	& $\sigma_s$	& 6.39	
			& \multirow{2}{*}{2e-4}	& $\sigma_b$	& 0.05	& 28.6		 \\
			&&$\sigma_{eff}$ & 3.35	&	& $\sigma_{eff}$& 0.05	& 15.0	 \\
		\hline
		\multirow{2}{*}{$\tau^+\tau^-$}	
			& \multirow{2}{*}{0.071}	& $\sigma_s$	& 2.01	
			& \multirow{2}{*}{0.034}	& $\sigma_b$	& 12.8	& 0.56	 \\
			&&$\sigma_{eff}$ & 1.58	&	& $\sigma_{eff}$& 4.97	& 0.71	 \\
		\hline
		\multirow{2}{*}{$\gamma\gamma$}	
			& \multirow{2}{*}{0.003}	& $\sigma_s$	& 0.077	
			& \multirow{2}{*}{---	}	& $\sigma_b$	& ---	& ---	 \\
			&&$\sigma_{eff}$ & ---	&	& $\sigma_{eff}$& ---	& ---	\\
		\hline
	\end{tabular}
	}
	\caption{Branching fractions, cross sections before and after cuts and $S/\sqrt{B}$ for the channels studied.}
\label{table:m-g-meas}
\end{table}

\begin{table}
	\centerline{
	\begin{tabular}{|c|r|l|l|l|}
		\hline
		\multirow{2}{*}{Channel}	&\multirow{2}{*}{}	& \multirow{2}{*}{$\Gamma_{H\rightarrow X} (MeV)$}	& \multirow{2}{*}{$\Delta M_H (MeV)$}	& \multirow{2}{*}{$Br(H^0\rightarrow X)$} \\
		&	&	&	&\\
		\hline
		\multirow{2}{*}{Total}	&
				Raw	& $4.56\pm1.52$	& $0.13\pm0.16$	&$0.96\pm0.04$ \\
			  &	Cut	& $5.57\pm1.33$	& $-0.02\pm0.14$& $0.65\pm0.01$ \\
		\hline
		\multirow{2}{*}{$b\bar{b}$}	&
				Raw	& $3.49\pm1.83$	& $-0.06\pm0.19$ & $0.67\pm0.05$ \\
		&		Cut	& $4.78\pm0.48$	& $0.01\pm0.05$	 & $0.271\pm0.001$ \\
		\hline
		\multirow{2}{*}{$WW^*$}	&
				Raw	& $4.06\pm0.24$	& $0.00\pm0.07$ & $0.217\pm0.001$ \\
			  &	Cut	& $3.96\pm0.17$	& $-0.16\pm0.04$ & $0.1271\pm0.0002$ \\
		\hline
		\multirow{2}{*}{$\tau^+\tau^-$}&
				Raw	& $4.82\pm4.46$	& $-0.54\pm0.47$ & $0.0623\pm0.0005$ \\
		&       Cut	& $0.84\pm2.97$	& $1.07\pm0.30$  & $0.24\pm0.23$ \\
		\hline
		\multirow{2}{*}{$\gamma\gamma$}&
				Raw	& $2.85\pm5.73$	& $-0.6\pm0.9$	& $0.0035\pm0.0001$ \\
		  &		Cut	& ---	& ---	& ---	\\
		\hline
	\end{tabular}
	}
	\caption{Fitted values of Higgs decay width, mass and branching ratio from simulated data. Mass values are the difference between the measured mass and the true mass of 126,000 MeV. Total integrated luminosity was $1~fb^{-1}$, or $71.4 pb^{-1}$ per data point.}
\label{table:h-fits}
\end{table}

\section{Higgs Measurements}	
In the previous section we fit simulated data to extract the properties of the Higgs. While it is clear that the $b\bar{b}$ and $WW^*$ channels will be the most useful for measuring Higgs properties, particularly with lower luminosities, the results of these fits are not reliable estimates of the achievable accuracy and precision of a muon collider. The values quoted were individual samples from trials that varied significantly in both accuracy and precision and which used the approximation that the background cross section, luminosity per point and beam resolution are well-known parameters. In this section we maintain this assumption and estimate the achievable accuracy and luminosity dependence of these measurements.

\subsection{Measurements With the $b\bar{b}$ Channel}
\begin{figure}
	\includegraphics[width=\textwidth]{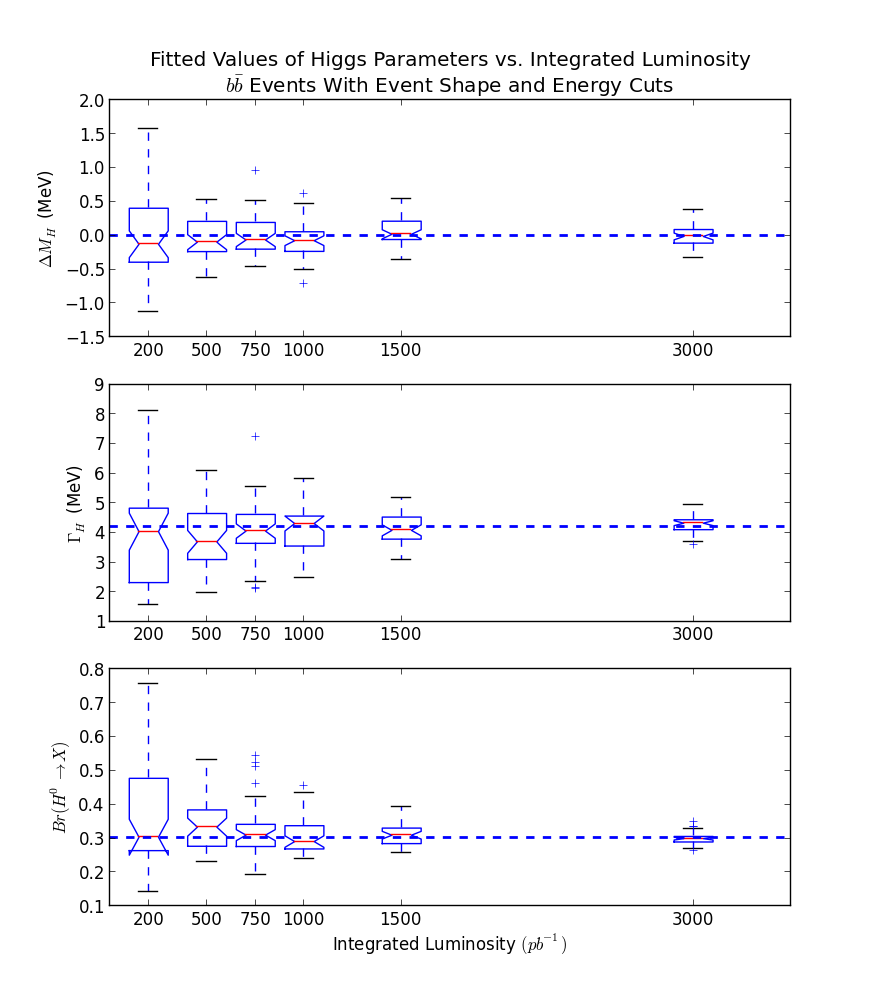}
	\caption{Box-and-whisker plots of fitted values of the Higgs mass, $b\bar{b}$ partial width and $b\bar{b}$ branching ratio for 40 experiments at each luminosity. Integrated luminosity is the total luminosity taken in 14 bins 4.2 MeV apart in a 60 MeV range centered on the Higgs mass. The boxes extend to the upper and lower quartiles of the data and the `whiskers' extend to the most extreme value within 1.5 times the inner-quartile range.\label{fig:fitted-vals-boxpl-bb}}
\end{figure}

The uncertainties in the measured values do not always reflect the accuracies of the measurements or their statistical variance from experiment to experiment. To get a better estimation we repeated the experiment of simulating $1~fb^{-1}$ of data and fitting it forty times. Figure~\ref{fig:fitted-vals-boxpl-bb} shows the results of this in box-and-whisker plots for a range of integrated luminosities. To reiterate, each experiment simulates taking data in a 60 MeV range around the Higgs peak with 14 bins separated by the beam width of 4.2 MeV. The integrated luminosity is the sum of luminosity taken in each bin. 

These plots demonstrate that our simplistic simulation and fitting experiment is on average accurate, but the statistical variance is high. While a more thorough analysis may provide more consistent results, we conclude here that at a given luminosity, the Higgs parameters can be measured to within the inner-quartile range given. At an integrated luminosity of $1~fb^{-1}$, we can use the $b\bar{b}$ channel with energy and event shape cuts to accurately measure the mass of the Higgs to within 0.3 MeV, the partial width to within 0.9 MeV and the branching ratio to within 0.09.

\subsection{Measurements with the $WW^*$ Channel}
\begin{figure}
	\includegraphics[width=\textwidth]{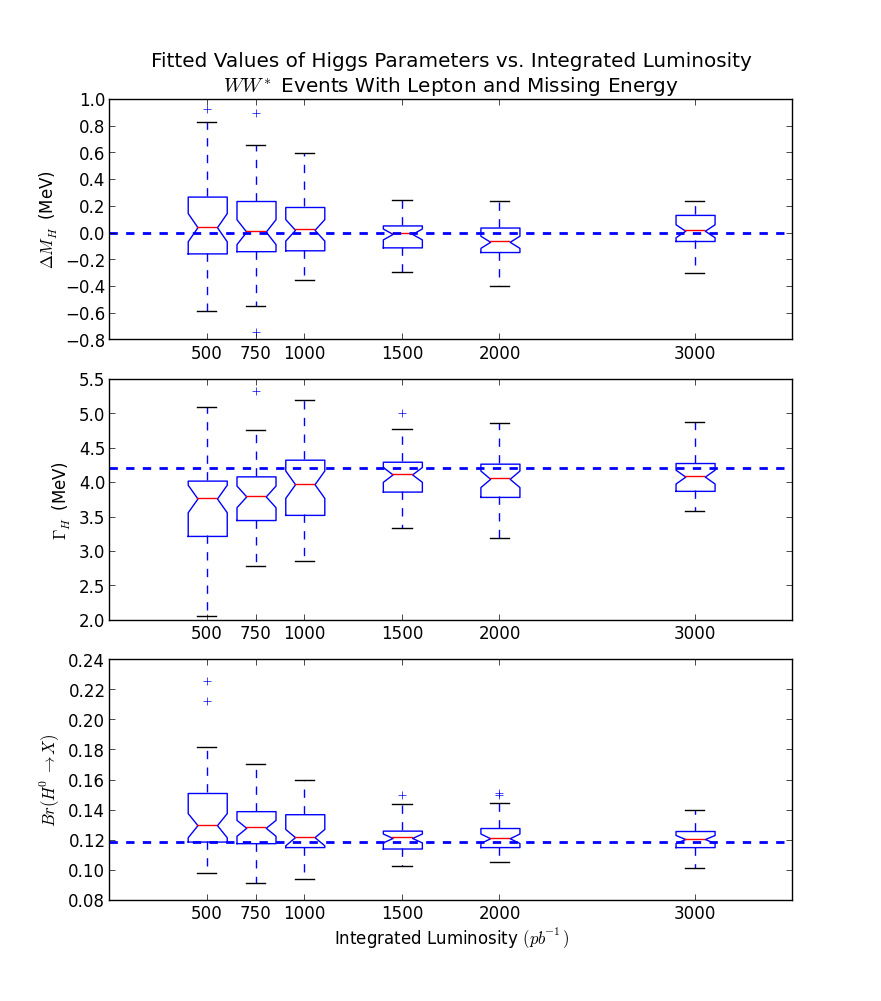}
	\caption{Box-and-whisker plots of fitted values of the Higgs mass, $WW^*$ partial width, and $WW^*$ branching ratio for 40 experiments at each luminosity. Integrated luminosity is the total luminosity taken in 14 bins 4.2 MeV apart in a 60 MeV range centered on the Higgs mass. The boxes extend to the upper and lower quartiles of the data and the `whiskers' extend to the most extreme value within 1.5 times the inner-quartile range.\label{fig:fitted-vals-boxpl-ww}}
\end{figure}

We performed the same simulated experiments using our estimated cross sections for the $WW^*$ channel and background, as shown in Figure~\ref{fig:fitted-vals-boxpl-ww}. We find that with an integrated luminosity of $1~fb^{-1}$, we can use the $WW^*$ channel with a lepton and missing energy to accurately measure the mass of the Higgs to within 0.38 MeV, the partial width to within 0.75 MeV and the branching ratio to within 0.02. These values can be found in Table~\ref{table:fitted-vals-boxpl}

\subsection{Combining Channels}
\begin{table}
	\begin{center}
		\begin{tabular}{|c|l|l|l|}
			\hline
			Channel	& $\delta M_H$ (MeV)	& $\delta \Gamma_H$ (MeV)	& $\delta Br(H^0\rightarrow X)$ \\ \hline
			$b\bar{b}$	& 0.30	& 0.60	& 0.09	\\ \hline
			$WW^*$		& 0.40	& 0.75	& 0.02	\\ \hline
			Combined	& 0.25	& 0.45	& ---	\\ \hline
		\end{tabular}
	\end{center}
	\caption{Accuracy of fitting parameters for simulated Higgs data. Values represent the inner quartile range (25\% to 75\%) of the values of 40 simulated experiments using $1~fb^{-1}$ total integrated luminosity. The combined values were calculated after each experiment using a weighted average.\label{table:fitted-vals-boxpl}}
\end{table}

\begin{figure}
	\includegraphics[width=\textwidth]{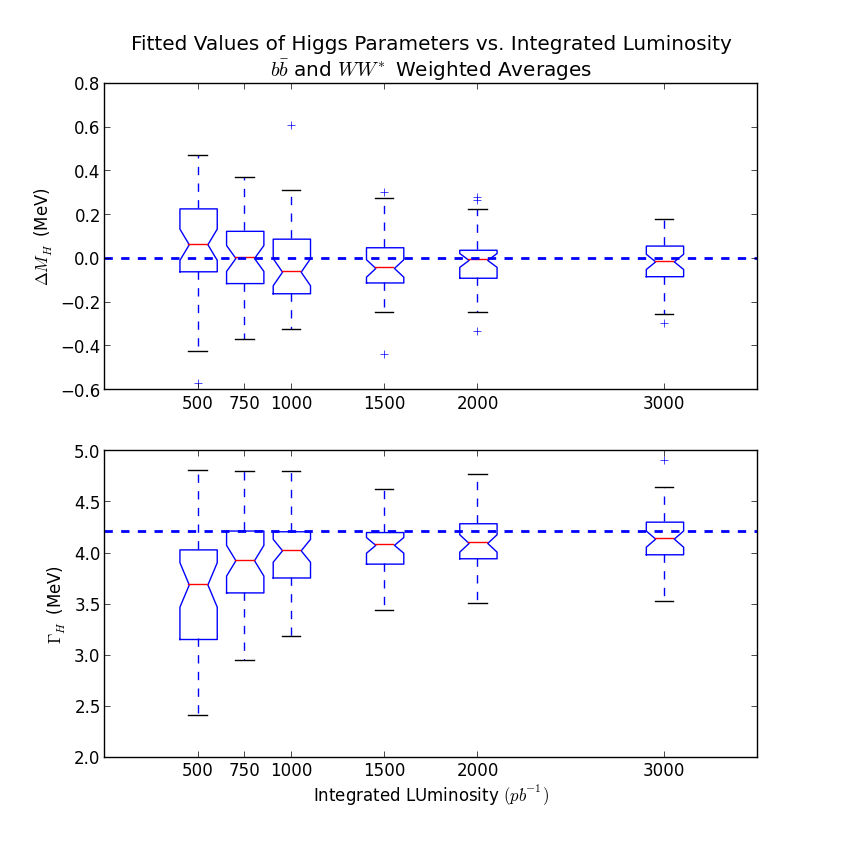}
	\caption{Box-and-whisker plots of fitted values of the Higgs mass and total width for 40 experiments at each luminosity. Integrated luminosity is the total luminosity taken in 14 bins 4.2 MeV apart in a 60 MeV range centered on the Higgs mass. The boxes extend to the upper and lower quartiles of the data and the `whiskers' extend to the most extreme value within 1.5 times the inner-quartile range.\label{fig:fitted-vals-boxpl-mix}}
\end{figure}

To measure the Higgs mass and total width more precisely, we took advantage of both channels. We did this by simulating data for both channels at the same time and taking their average, weighted by the uncertainty in the fits. For example, the formula used for the width was:
\begin{equation}
	\delta \overline{\Gamma_H} = \frac{\delta \Gamma_{b\bar{b}}}{\delta \Gamma_{b\bar{b}} + \delta \Gamma_{WW^*}}\Gamma_{WW^*} + \frac{\delta \Gamma_{WW*}}{\delta \Gamma_{b\bar{b}} + \delta \Gamma_{WW^*}}\Gamma_{b\bar{b}}\label{eq:weight-avg}
\end{equation}
As shown in Figure~\ref{fig:fitted-vals-boxpl-mix}, the mass measurement was found to be accurate within 0.25 MeV and the total width was accurate within 0.45 MeV. All the estimated accuracies can be found in Table~\ref{table:fitted-vals-boxpl}.

\section{Finding the Higgs Peak}
	Before precise measurements of the Higgs can be made it will be necessary to search for the Higgs by scanning the beam energy and looking for a significant signal. Because s-channel resonant production only happens with the beam center-of-mass energy directly on the Higgs mass, this is not a trivial matter and it is important to know how much luminosity will be required to find the Higgs, what channels will be most useful, and what beam momentum spread will be ideal; too wide and the effective Higgs cross section will be miniscule, too narrow and it will require too many data points.

	To answer this question we devised a simple scanning scheme and calculated how much luminosity would be needed to guarantee finding a significant Higgs signal. We begin by assuming the Higgs mass has been measured at 126.0 GeV to within 100 MeV, based on estimates of the potential resolution possible at the LHC.\cite{lhc-higgs-width} We then calculate how much luminosity will be needed at each point to guarantee, to a given confidence level, a statisitcally significant signal assuming the beam center of mass energy is exactly at the Higgs mass. Then we conduct a search guided by the existing mass measurement, taking data at energies calculated to add the most probability to finding the Higgs.

	\subsection{Guaranteeing a Significant Signal}
	The first step is finding out how much luminosity will be needed to guarantee a significant signal if we are taking data on the peak. We will assume we are requiring a signal with $5\sigma$ significance. To do this, we first calculate $N$, the number of events that must be observed to count as a significant signal, given a certain luminosity and background. We use the Poissonian distribution with an expected value of $\mathcal{L}\times \sigma_b$ events to calculate the probability $p$ of a given signal and require an $N$ such that $p$ is less than $5\sigma$ where $\sigma_b$ is the background cross section.
	\begin{equation}
		p = \int_N^\infty \Pr(X=n|\mathcal{L}\times \sigma_b)\ dn\label{eq:p-value}
	\end{equation}
	Then we calculate the confidence level $\alpha$, the probability of not seeing more than $N$ events given we are taking data on the peak, where the expected number of events is $\mathcal{L}\times(\sigma_b+\sigma_s)$ and $\sigma_s$ is the signal cross section.
	\begin{equation}
		\alpha = \int_0^N \Pr(X=n|\mathcal{L}\times(\sigma_b+\sigma_s))\ dn\label{eq:alpha}
	\end{equation}
	With the Poissonian, these integrals reduce to a sum. We calculate the luminosity required for the confidence level $\alpha$ to reach a range of values ($1\sigma$, $1.64\sigma$, $2\sigma$, $2.58\sigma$, $3\sigma$, $4\sigma$, $5\sigma$).

	\subsection{Search Strategy}
	We can now evaluate the probability of finding a significant Higgs signal at a given energy and luminosity based on the confidence level $\alpha$ provided by that luminosity and a Gaussian probability distribution based on the LHC measurements. We then order the search by maximizing the probability of finding a significant signal at each step, taking enough data at each point to raise $\alpha$ to its next value. Data points are separated in energy by the width of the beam spread Gaussian to minimize the risk of missing the Higgs peak between data points while also minimizing the number of data points needed, since sums of Gaussians separated by their standard deviation form an approximately flat distribution. We then calculated how much total integrated luminosity would be required for this search strategy to guarantee finding the Higgs peak at a range of luminosities, given a certain signal and background and using a range of beam widths.

	\subsection{Required Luminosity}
	\begin{figure}
		\includegraphics[width=\textwidth]{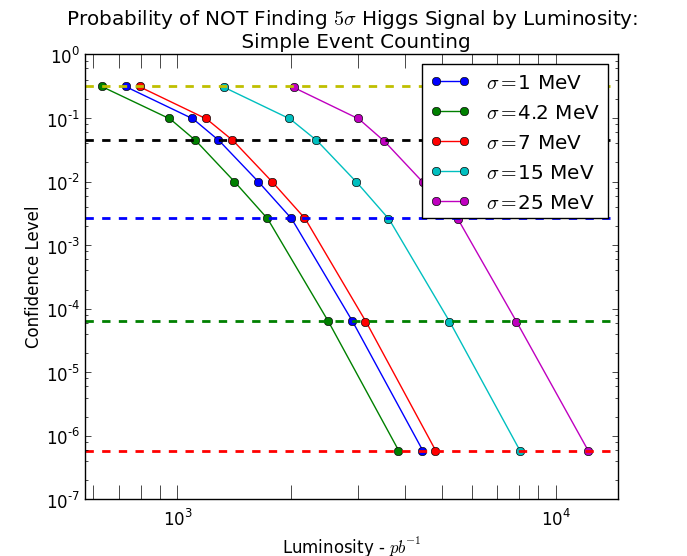}
		\caption{Probability of not finding a significant Higgs signal using our search strategy as a function of luminosity for a range of beam energy resolutions. Calculated for a search using simple event counting only.\label{fig:lum-needed-total-raw}}
	\end{figure}

\begin{figure}
		\includegraphics[width=\textwidth]{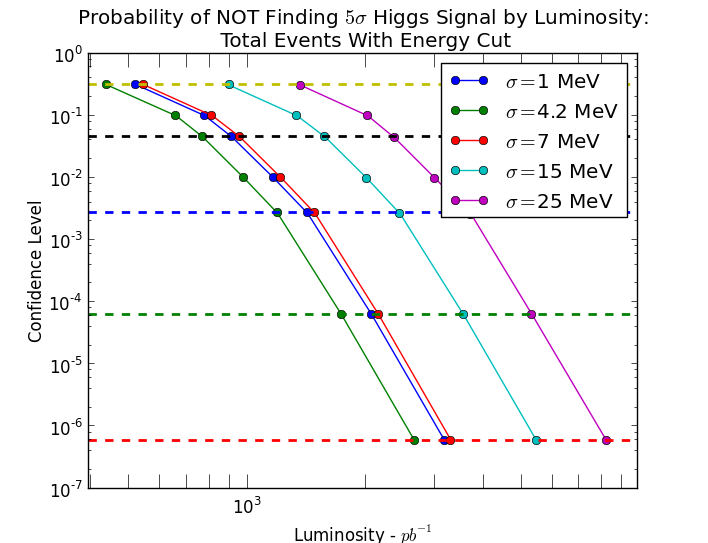}
		\caption{Probability of not finding a significant Higgs signal using our search strategy as a function of luminosity for a range of beam energy resolutions. Calculated for a search using simple event counting and requiring events with total visible energy $E_{tot} > 98 GeV$.\label{fig:lum-needed-total-cut}}
	\end{figure}

	First, we looked at the results of using simple event counting, using all events except for $\mu^+\mu^- \rightarrow Z^0 \rightarrow \nu_{\ell} \bar{\nu_\ell}$. Figure~\ref{fig:lum-needed-total-raw} makes it clear that total event counting would not be practical; it would require on the order of inverse femtobarns of integrated luminosity to guarantee finding the Higgs at a $3\sigma$ confidence level. It does however reveal that a beam width near the actual Higgs width would offer the ideal balance. We then looked at the improvement that could be achieved with the energy cuts described above (See Figure~\ref{fig:lum-needed-total-cut}). Required luminosities for these and the following channels are listed in Table~\ref{table:lum-needed}. Additional plots of luinosity can be found in Appendix~\ref{sec:lum-needed}.

	We next tried the search strategy using tagging of $b\bar{b}$ events and the effect of implementing the energy and event shape cuts described above. As shown in Figures~\ref{fig:lum-needed-bbar-raw} and~\ref{fig:lum-needed-bbar-cut}, the cuts reduced the required luminosity by a factor of about two. As expected, a beam width near the actual Higgs width continues to be the most effective.

	Finally, we looked at $H^0 \rightarrow WW^*$ events with a lepton and missing energy, as described above. The high purity and relatively large cross section of this channel makes it significantly better than other channels for searching for the Higgs (Figure~\ref{fig:lum-needed-ww-cut}). Interestingly, it also appears that the low background removes the necessity for a beam with a width near the Higgs decay width; beam widths in the range from 4.2 MeV to 25 MeV fare similarly with this search strategy in terms of luminosity.

	\subsection{Combined Channels}

	\begin{figure}
		\includegraphics[width=\textwidth]{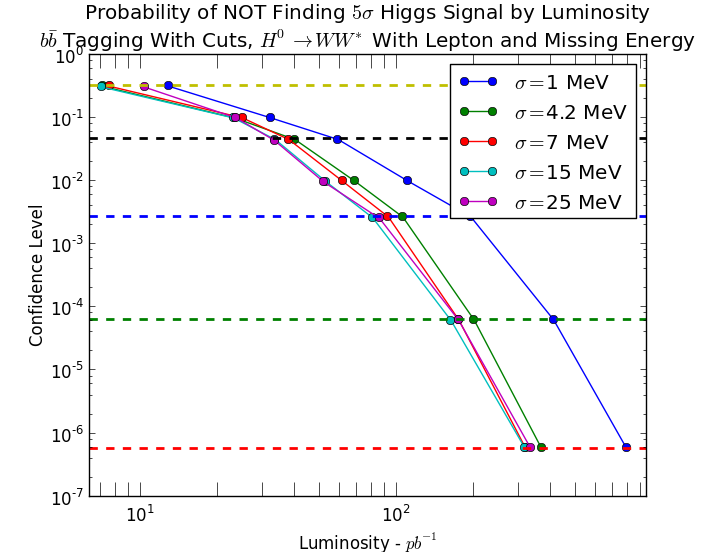}
		\caption{Probability of not finding a significant Higgs signal using our search strategy as a function of luminosity for a range of beam energy resolutions. Calculated for a search combining the $b\bar{b}$ and $WW^*$ channels, using the cuts described above.\label{fig:lum-needed-ww-bb}}
	\end{figure}

	Fortunately, there is no need to use only one channel or another when searching for the Higgs, so next we examine the effects of using both the $b\bar{b}$ and $WW^*$ channels in our search strategy. As seen in Figure~\ref{fig:lum-needed-ww-bb}, this offers improvement over using any one channel alone. This is because the probability of the background fluctuating significantly in two channels at once is equal to the product of the probabilities for each channel fluctuating on their own. Again we see that the beam resolution makes little difference and while wider beam spreads seem to require slightly less luminosity, narrower beams will measure the Higgs peak with greater precision, reducing luminosity needed for more precise measurements. 

\begin{table}
	\begin{tabular}{|c|l|r|r|}
		\hline
		\multirow{2}{*}{Channel}	& $\sigma_{sig}\ (pb)$	& \multicolumn{2}{|c|}{Luminosity Required $(pb^{-1})$} \\ \cline{3-4}
		& $\sigma_{bkgr}\ (pb)$	& $CL = 3\sigma$ & $CL = 5\sigma$ \\ \hline
		\multirow{2}{*}{Total} & $\sigma_s = 28.3$ & 
			\multirow{2}{*}{1,723}	& \multirow{2}{*}{3,840} \\
			& $\sigma_b = 301.4$	&	& \\ \hline
		\multirow{2}{*}{Total (Cut)} & $\sigma_s = 22.4$ & 
			\multirow{2}{*}{1,193}	& \multirow{2}{*}{2,666} \\
			& $\sigma_b = 126.4$	&	& \\ \hline
		\multirow{2}{*}{$b\bar{b}$} & $\sigma_s = 16.5$ & 
			\multirow{2}{*}{1,033}	& \multirow{2}{*}{2,317} \\
			& $\sigma_b= 57.2$	&	& \\ \hline
		\multirow{2}{*}{$b\bar{b}$ (Cut)} & $\sigma_s = 8.64$ & 
			\multirow{2}{*}{697}	& \multirow{2}{*}{1,593} \\
			& $\sigma_b= 8.45$	&	& \\ \hline
		\multirow{2}{*}{$WW^*$} & $\sigma_s = 6.39$ & 
			\multirow{2}{*}{146}	& \multirow{2}{*}{389} \\
			& $\sigma_b= 0.05$	&	& \\ \hline
		\multirow{2}{*}{$WW^*$ (Cut)} & $\sigma_s = 3.35$ & 
			\multirow{2}{*}{325}	& \multirow{2}{*}{812} \\
			& $\sigma_b= 0.05$	&	& \\ \hline
		\multirow{2}{*}{$b\bar{b},\ WW^*$} & $\sigma_s = --$ & 
			\multirow{2}{*}{105}	& \multirow{2}{*}{368} \\
			& $\sigma_b= --$	&	& \\ \hline
	\end{tabular}
	\caption{Required luminosity to guarantee finding a $5\sigma$ Higgs signal with confidence level $\alpha=3\sigma,5\sigma$.\label{table:lum-needed}} 
\end{table}

\section{Discussion and Conclusion}
The Higgs boson is a particle of fundamental importance to physics and measuring its properties with precision will allow us to probe the limits of the Standard Model and may point the way towards non-Standard model physics. Using simple estimates of physics backgrounds and separable signal we have estimated that with $1~fb^{-1}$ of integrated luminosity a hypothetical muon collider Higgs factory operating at the Higgs s-channel resonance could measure the mass of a Standard Model 126 GeV Higgs to within 0.25 MeV and its total width to within 0.45 MeV. We estimated that with a beam spread of 4.2 MeV, approximately 368 $pb^{-1}$ total integrated luminosity would be required to guarantee locating the narrow Higgs peak. We believe that these preliminary results strongly motivate further research and development towards the construction of a muon collider Higgs factory.

Our estimations assume that there is no machine-induced background and that the detector has excellent tracking and calorimetry. Our results demonstrate the value of the high Higgs cross section and narrow beam energy spread available at a muon collider. These two factors enable the direct measurement of the Higgs mass and width by scanning the Higgs s-channel resonance, which is not possible at any $e^+e^-$ collider. Our study of the physics-induced background and separation of the Higgs signal showed that significant reduction of the physics background can be achieved by a detector with high energy and spatial resolution. We believe that this report justifies more in-depth analysis of Higgs channels and their backgrounds, for example the reconstruction of $H^0\rightarrow WW^*\rightarrow 4j$ events using learning algorithms or the application of flavor-tagging techniques to tag $b\bar{b}$ events. 

Machine-induced backgrounds, mainly from muon decays in the beam, present an additional difficulty which has not yet been studied in great detail. We believe that in addition to significant shielding in the detector cone and endcaps, it may be important to have a calorimeter with high spatial and temporal resolution. Our results motivate an in-depth analysis of the machine-induced background including simulation in a highly segmented, totally-active, dual readout calorimeter such as the MCDRCal01 detector concept.

\section{Acknowledgements}
We would like to thank Young-Kee Kim, Estia Eichten, Ron Lipton, Nikolai Terentiev, Norman Graf and Steve Mrenna for support, discussions and collaboration. 

\appendix
\section{Additional Figures}
\subsection{Simulated Event Counts}
\label{sec:sim-evt-cts}

\begin{figure}[H]
	\includegraphics[width=0.6\textwidth]{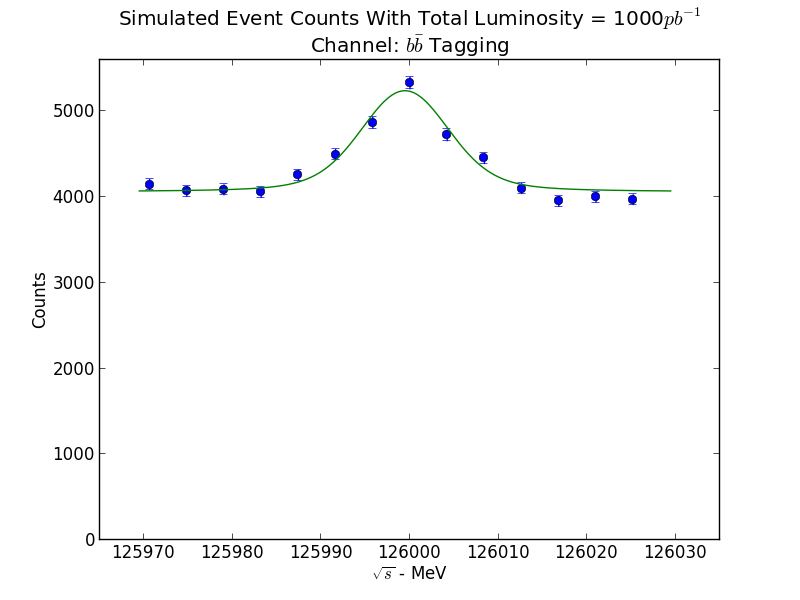}
	\caption{Simulated event counts for a scan across a 126.0 GeV Higgs peak with a 4.2 MeV wide Gaussian beam spread, counting all $X\rightarrow b\bar{b}$ events. Data is taken in a 60 MeV range centered on the Higgs mass in bins separated by the beam width of 4.2 MeV. Event counts are calculated as Poisson-distributed random variables and the data is fit to a Breit-Wigner convoluted with a Gaussian plus linear background. The fit width is $3.49\pm1.83$ MeV, the error in the mass measurement is $-0.06\pm0.19$ MeV and the branching ratio is measured at $0.67\pm0.05$. Total luminosity is $1000pb^{-1}$, or $71.4 pb^{-1}$ per point.}
\label{data-fit-bbbar-raw}
\end{figure}

\begin{figure}[H]
	\includegraphics[width=0.6\textwidth]{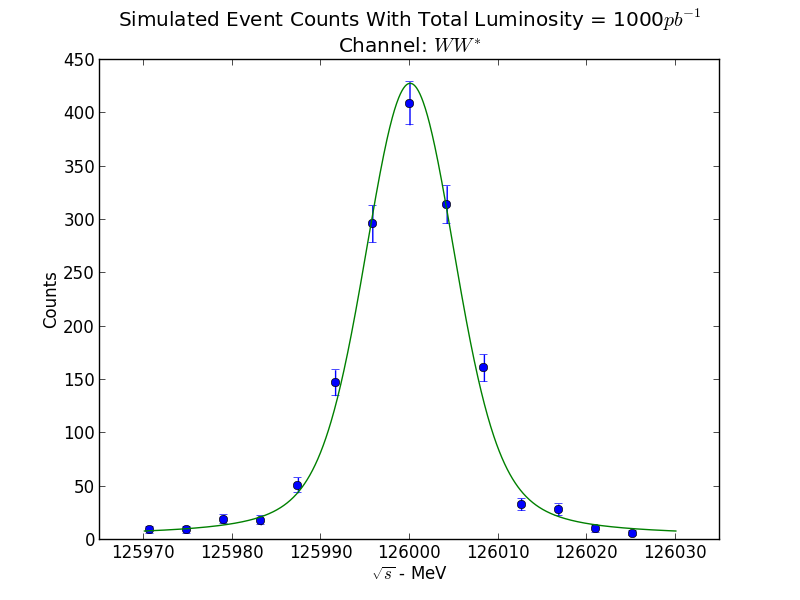}
	\caption{Simulated event counts for a scan across a 126.0 GeV Higgs peak with a 4.2 MeV wide Gaussian beam spread, counting all $H^0\rightarrow WW^*$ events with a minimal background. Data is taken in a 60 MeV range centered on the Higgs mass in bins separated by the beam width of 4.2 MeV. Event counts are calculated as Poisson-distributed random variables and the data is fit to a Breit-Wigner convoluted with a Gaussian plus linear background. The fit width is $4.06\pm0.24$ MeV, the error in the mass measurement is $0.00\pm0.07$ MeV and the branching ratio is measured at $0.217\pm0.001$. Total luminosity is $1000pb^{-1}$, or $71.4 pb^{-1}$ per point.}
\label{data-fit-ww-raw}
\end{figure}

\begin{figure}[H]
	\includegraphics[width=0.6\textwidth]{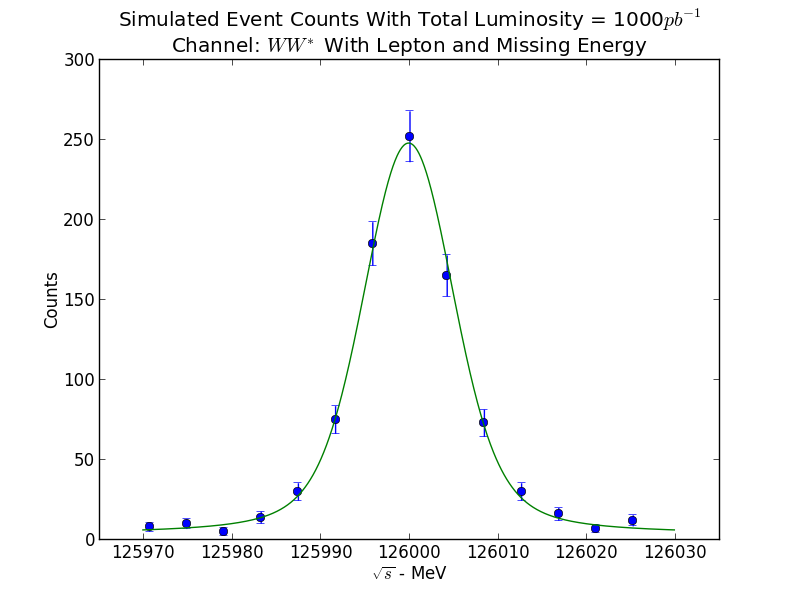}
	\caption{Simulated event counts for a scan across a 126.0 GeV Higgs peak with a 4.2 MeV wide Gaussian beam spread, counting all $H^0\rightarrow WW^*\rightarrow$ lepton + missing energy events with a minimal background. Data is taken in a 60 MeV range centered on the Higgs mass in bins separated by the beam width of 4.2 MeV. Event counts are calculated as Poisson-distributed random variables and the data is fit to a Breit-Wigner convoluted with a Gaussian plus linear background. The fit width is $3.96\pm0.17$ MeV, the error in the mass measurement is $-0.16\pm0.04$ MeV and the branching ratio is measured at $0.1271\pm0.0002$. Total luminosity is $1000pb^{-1}$, or $71.4 pb^{-1}$ per point.}
\label{data-fit-ww-cut}
\end{figure}

\begin{figure}[H]
	\includegraphics[width=0.6\textwidth]{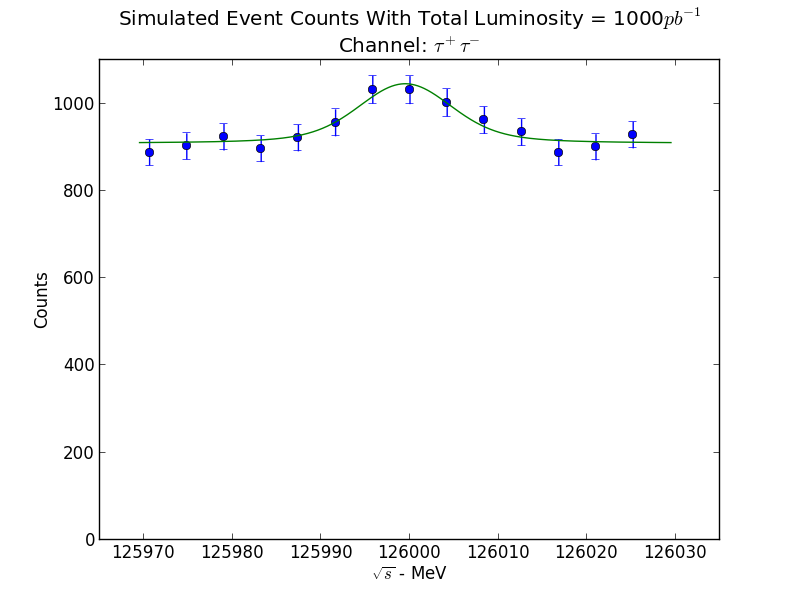}
	\caption{Simulated event counts for a scan across a 126.0 GeV Higgs peak with a 4.2 MeV wide Gaussian beam spread, counting all $X\rightarrow \tau^+\tau^-$ events. Data is taken in a 60 MeV range centered on the Higgs mass in bins separated by the beam width of 4.2 MeV. Event counts are calculated as Poisson-distributed random variables and the data is fit to a Breit-Wigner convoluted with a Gaussian plus linear background. The fit width is $4.82\pm4.46$ MeV, the error in the mass measurement is $-0.54\pm0.47$ MeV and the branching ratio is measured at $0.0623\pm0.0005$. Total luminosity is $1000pb^{-1}$, or $71.4 pb^{-1}$ per point.}
\label{data-fit-tt-raw}
\end{figure}

\begin{figure}[H]
	\includegraphics[width=0.6\textwidth]{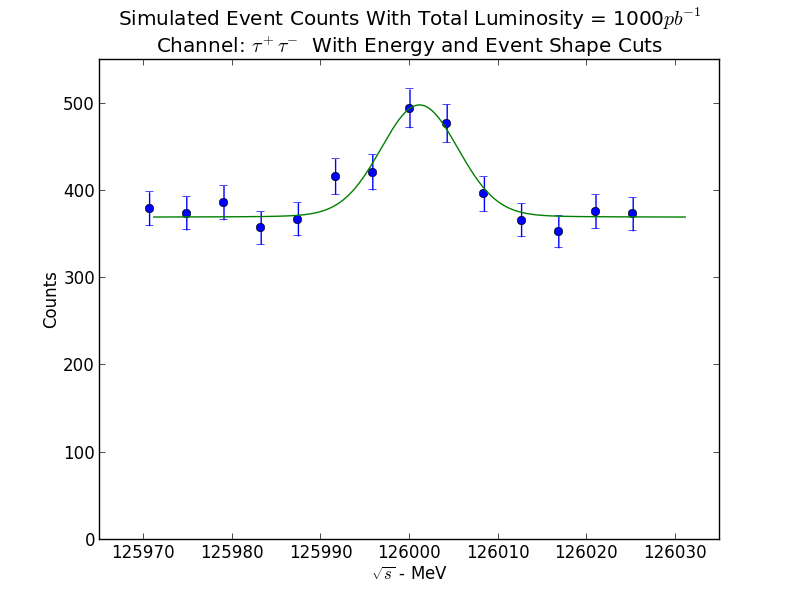}
	\caption{Simulated event counts for a scan across a 126.0 GeV Higgs peak with a 4.2 MeV wide Gaussian beam spread, counting $X\rightarrow \tau^+\tau^-$ events with a total energy of at least 60.0 GeV visible to the detector and cutting on event shape parameters. Data is taken in a 60 MeV range centered on the Higgs mass in bins separated by the beam width of 4.2 MeV. Event counts are calculated as Poisson-distributed random variables and the data is fit to a Breit-Wigner convoluted with a Gaussian plus linear background. The fit width is $0.84\pm2.97$ MeV, the error in the mass measurement is $1.07\pm0.30$ MeV and the branching ratio is measured at $0.24\pm0.23$. Total luminosity is $1000pb^{-1}$, or $71.4 pb^{-1}$ per point.}
\label{data-fit-tt-cut}
\end{figure}

\begin{figure}[H]
	\includegraphics[width=0.6\textwidth]{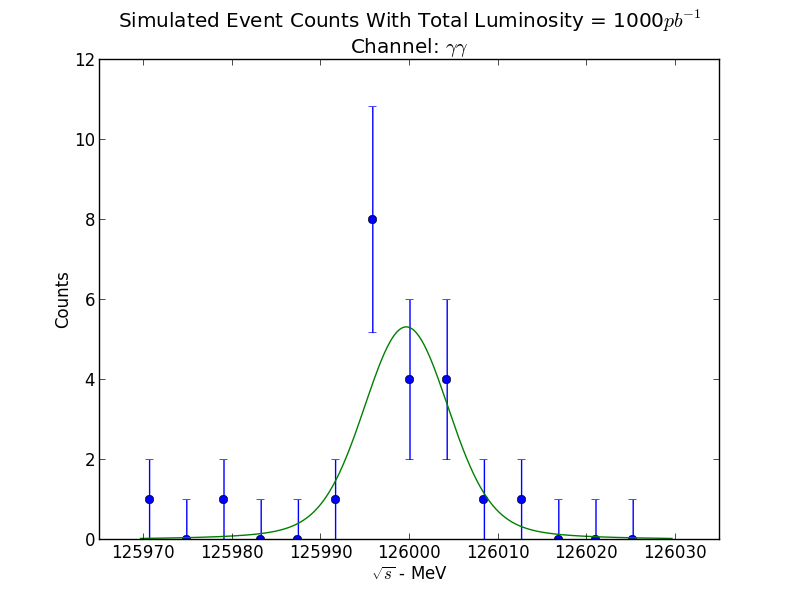}
	\caption{Simulated event counts for a scan across a 126.0 GeV Higgs peak with a 4.2 MeV wide Gaussian beam spread, counting all $H^0\rightarrow \gamma\gamma$ events. Data is taken in a 60 MeV range centered on the Higgs mass in bins separated by the beam width of 4.2 MeV. Event counts are calculated as Poisson-distributed random variables and the data is fit to a Breit-Wigner convoluted with a Gaussian plus linear background. The fit width is $2.85\pm5.73$ MeV, the error in the mass measurement is $-0.6\pm0.9$ MeV and the branching ratio is measured at $0.0035\pm0.0001$. Total luminosity is $1000pb^{-1}$, or $71.4 pb^{-1}$ per point.}
\label{data-fit-gg-cut}
\end{figure}

\subsection{Search Strategy}
\label{sec:lum-needed}

\begin{figure}[H]
		\includegraphics[width=0.8\textwidth]{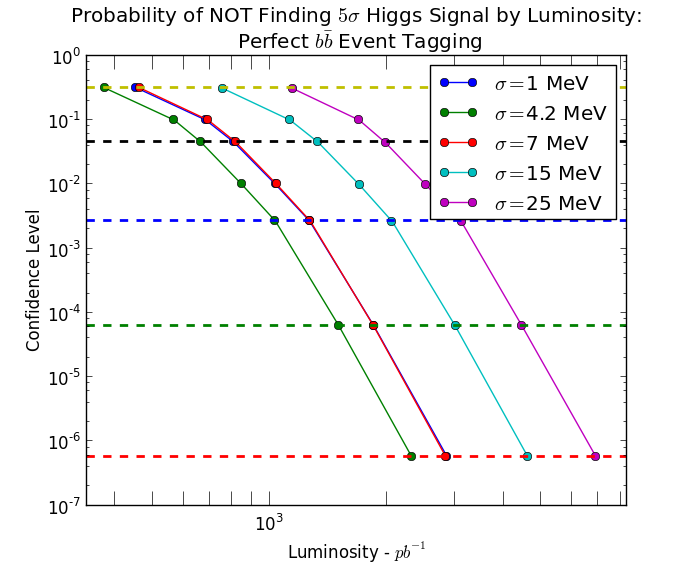}
		\caption{Probability of not finding a significant Higgs signal using our search strategy as a function of luminosity for a range of beam energy resolutions. Calculated for a search using perfect $b\bar{b}$ tagging.\label{fig:lum-needed-bbar-raw}}
	\end{figure}

	\begin{figure}[H]
		\includegraphics[width=0.8\textwidth]{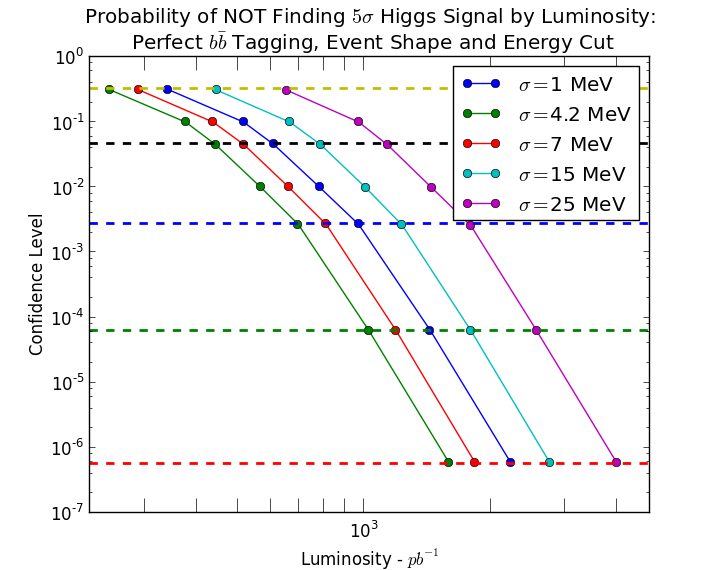}
		\caption{Probability of not finding a significant Higgs signal using our search strategy as a function of luminosity for a range of beam energy resolutions. Calculated for a search using perfect $b\bar{b}$ tagging and event shape and energy cuts.\label{fig:lum-needed-bbar-cut}}
	\end{figure}

	\begin{figure}[H]
		\includegraphics[width=0.8\textwidth]{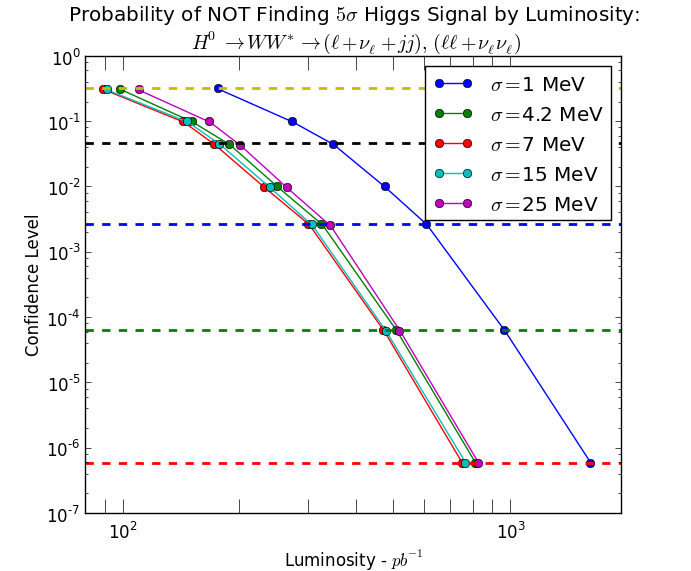}
		\caption{Probability of not finding a significant Higgs signal using our search strategy as a function of luminosity for a range of beam energy resolutions. Calculated for a search using $H^0\rightarrow WW^*\rightarrow$ lepton + missing energy events with minimal background.\label{fig:lum-needed-ww-cut}}
	\end{figure}

\subsection{Event Displays}
\label{sec:evt-display}
The following event displays were generated using the WIRED4 plug-in for JAS3 using $\mu^+\mu^-\rightarrow H^0$ events with $\sqrt{\hat{s}} = M_H = 125.0 GeV$ generated by PYTHIA 6.4 and simulated in the MCDRCal01 concept detector using SLIC.\@

\begin{figure}[h]
	\includegraphics[width=\textwidth]{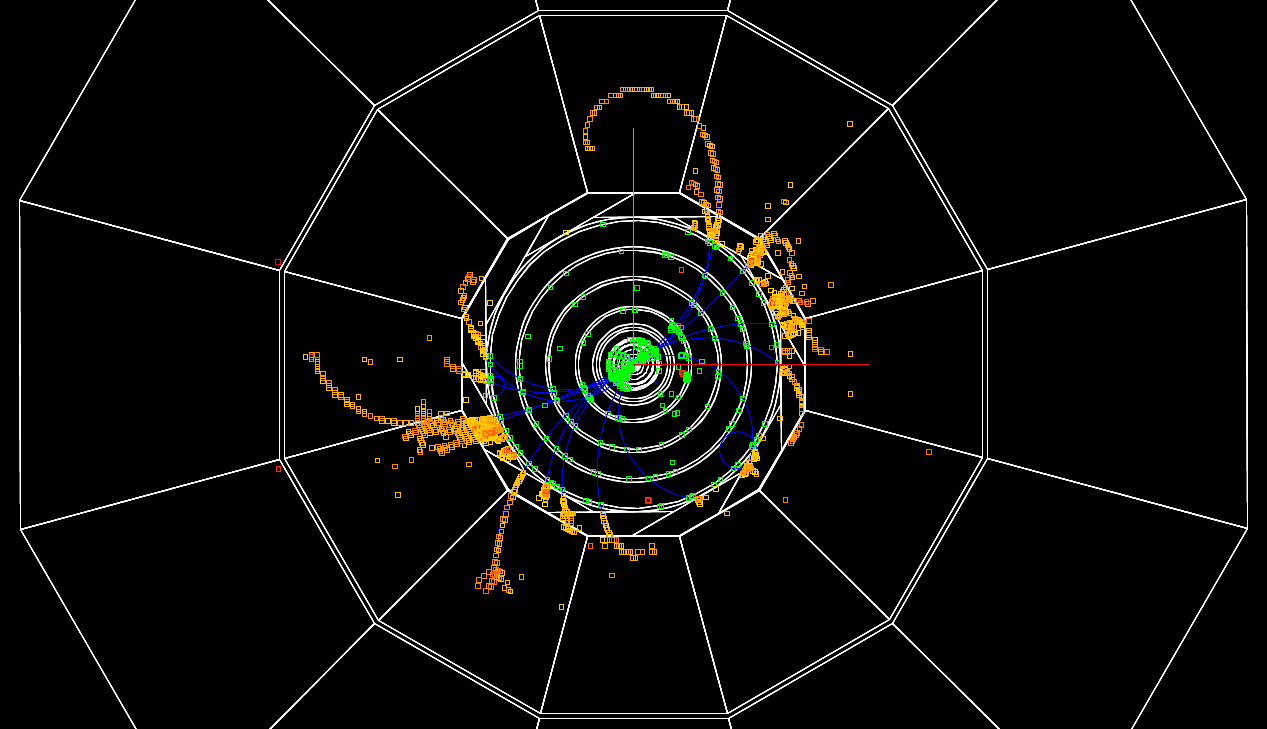}
	\caption{Example $H^0\rightarrow b\bar{b}$ event display.\label{fig:evt-disp-bbbar}}
\end{figure}

\begin{figure}
	\includegraphics[width=\textwidth]{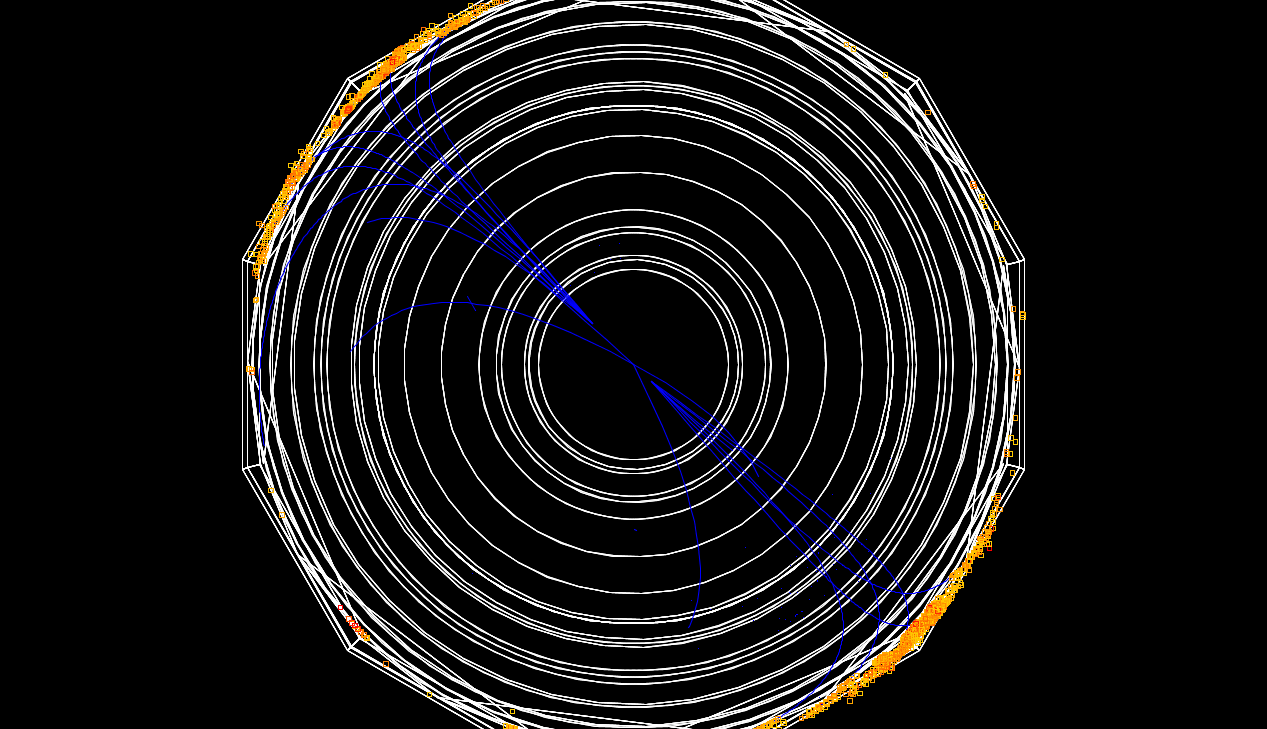}
	\caption{Example $H^0\rightarrow b\bar{b}$ event display with fisheye projection to illustrate the displaced b vertices.\label{fig:evt-disp-bbbar-fish}}
\end{figure}

\begin{figure}
	\includegraphics[width=\textwidth]{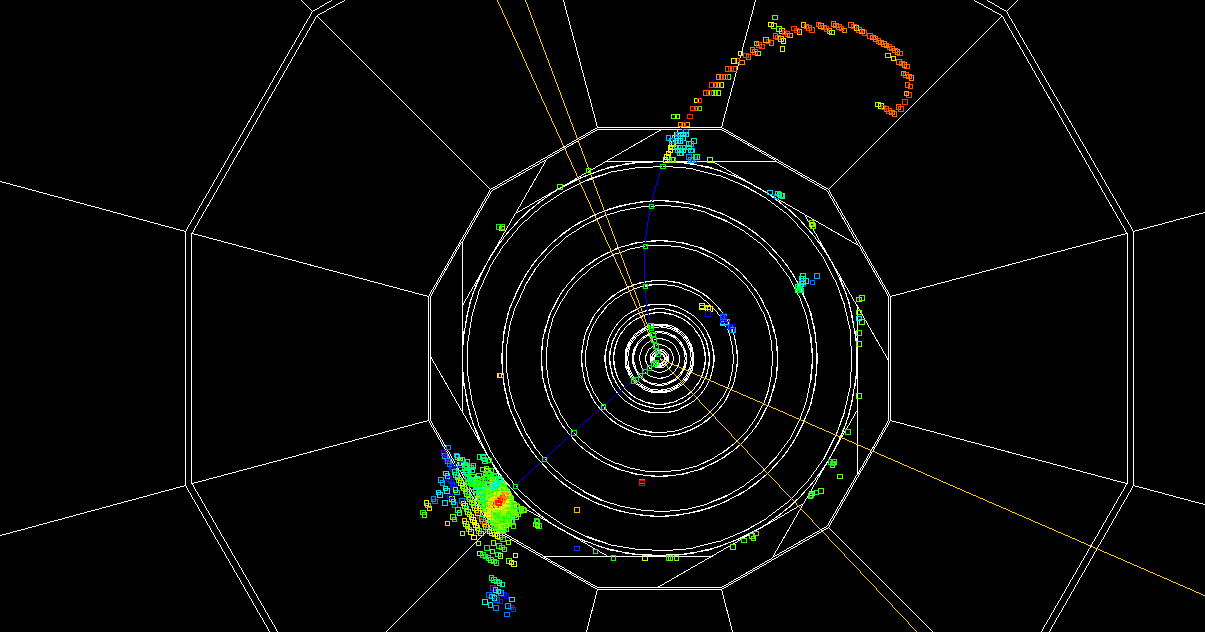}
	\caption{Example $H^0\rightarrow WW^*\rightarrow e^- + \bar{\nu_e} + \tau^+ + \nu_\tau$ event display. Yellow lines indicate neutrinos.\label{fig:evt-disp-ww-e-tau-nunu}}
\end{figure}

\begin{figure}
	\includegraphics[width=\textwidth]{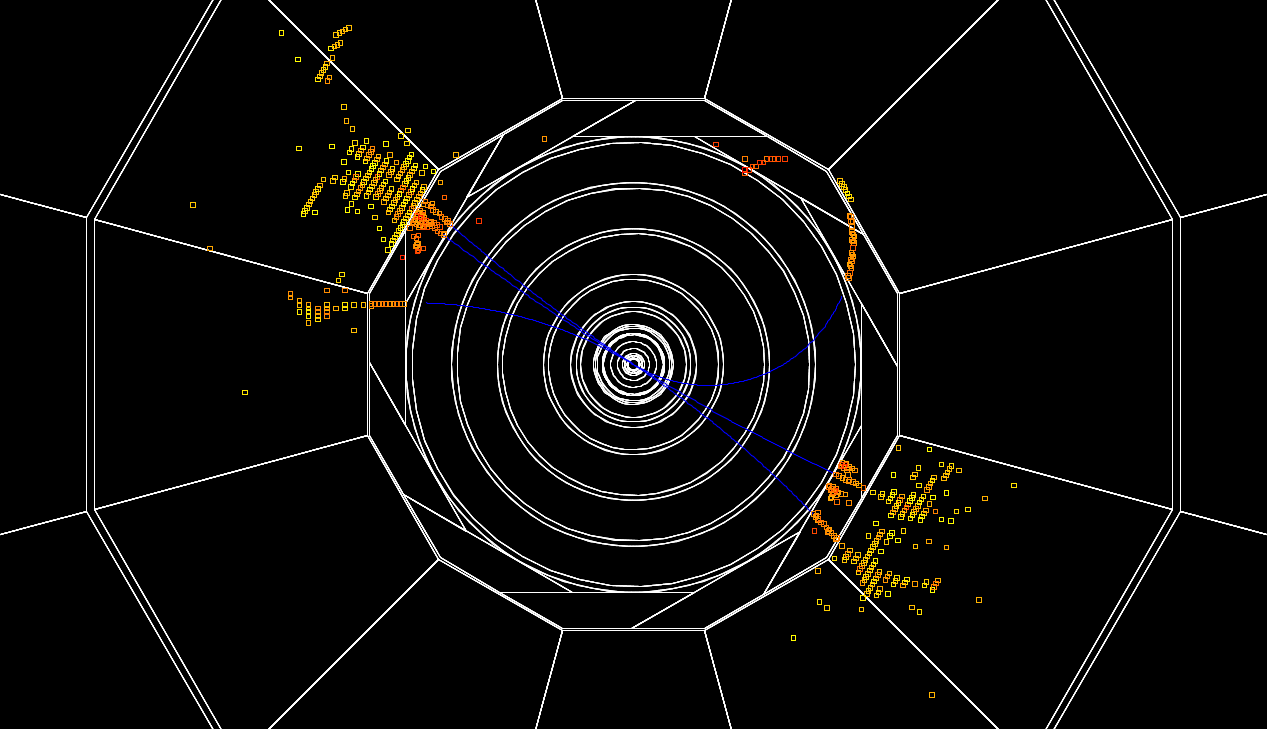}
	\caption{Example $H^0\rightarrow \tau^+\tau^-$ event display.\label{fig:evt-disp-tautau}}
\end{figure}

\bibliographystyle{plain}
\bibliography{bib/report}
\end{document}